# Parallel multifocus optical-resolution photoacoustic microscopy based on tunable acoustic gradient lens and fiber delay network

## XIANLIN SONG

*School of Information Engineering, Nanchang University, Nanchang, China*

*songxianlin@ncu.edu.cn*

**We developed a multifocus optical-resolution photoacoustic microscope (MPAM) with extended depth of field (DoF) by using a tunable acoustic gradient (TAG) lens and a pulse delay method. TAG lens is designed to control the divergence angle of the laser beam. Three fibers with different length are used to output three laser pulses with different delay time. By controlling the fire time of the laser pulses, we can get three focal spot through TAG lens in our system in one A-line data acquisition. The DoF of MPAM measured using a vertically tilted carbon fiber is about 775 μm. The lateral resolution is estimated to be 2.8 μm. A phantom consists of tungsten wires and a zebrafish is used to verify the large depth-of-field of our system.**

Photoacoustic imaging is a high-resolution and high-contrast technique, which combines optical contrast with ultrasonic detection to map the distribution of the absorbing pigments in biological tissues [1-5]. It can provide structural, functional and molecular imaging of tissues [6-8]. Photoacoustic microscopy (PAM) has also been developed as a promising tool for tumor detection[9], monitoring of blood oxygenation[10]. It can be classified into two categories: optical-resolution (OR-) and acoustic-resolution (AR-) PAM [11, 12]. In the OR-PAM, the lateral resolution is determined by optical focusing. However, for most OR-PAM systems, the DoF is quite limited, due to a single depth of focus, only a narrow depth range in focus can be acquired. The small DoF will prevent OR-PAM to achieve dynamic information in depth direction.

Motorized stage is widely used to improve depth of field (DoF) [13, 14], but the focus-shifting speed is quite limited by mechanical inertia. The chromatic aberration has been exploited previously in PAM [15], the system can generate multi-focus along the depth direction, but sacrifices the capability of functional imaging. Double illumination PAM system illuminating the sample from both top and bottom sides simultaneously to improve DoF, but it is only valid in transmission-mode OR-PAM [16]. Replacing the diffracting Gaussian beam with a non-diffracting Bessel beam, large DoF of 1 mm with a lateral resolution of 7 μm have been achieved [17]. However, a non-linear method must be used to suppress the artifacts introduced by the side lobes of the Bessel beam, which increases the complexity of the imaging process. Electrically tunable lens (ETL) has also been introduced in OR-PAM [15], the focus-shifting time is about 15 ms. It is fast enough for pulsed lasers with a repetition rate of 20 Hz, but insufficient for lasers with repetition rates of many kHz [18].

In this letter, we report a novel OR-PAM that can achieve large DoF. The report system employs a high-speed TAG lens and a fiber delay network to achieve three focuses. The DoF is estimated by a vertically tilted carbon fiber. A tungsten wire network and a zebrafish were also imaged to demonstrate the feasibility of our system.

The multifocus photoacoustic microscope referred to as “MF-PAM” hereafter, is illustrated in Fig. 1. The system is in transmission mode. An Nd: YLF laser (IS8II-E, EdgeWave GmbH) irradiate laser light at the wavelength of 523 nm and a pulse repetition rate of 1 KHz. The laser beam is reshaped by an iris with a diameter of 0.8 mm and then expanded by a pair of plano-convex lenses (L2， L3). A 50 μm-diameter pinhole (PH) is placed in the expander as a spatial filter. Then the laser beam was split into three beams, and the three beams were coupled into three multimode fibers (core diameter of 50 μm), respectively. The lengths of three fibers are 1 m, 26 m, 51 m, respectively. The lights come out of three multimode fibers are coupled into a single mode fiber together. The distal end of the single mode fiber is inserted into a fiber port (PAFA-X-4-A, Thorlabs) which transforms the guided laser beam into collimated beam with an output $1/e^2$ waist diameter of 0.65 mm. Then the optical axis of the laser beam is turned to vertical to reduce the gravity induced Y-coma aberration of the TAG lens.

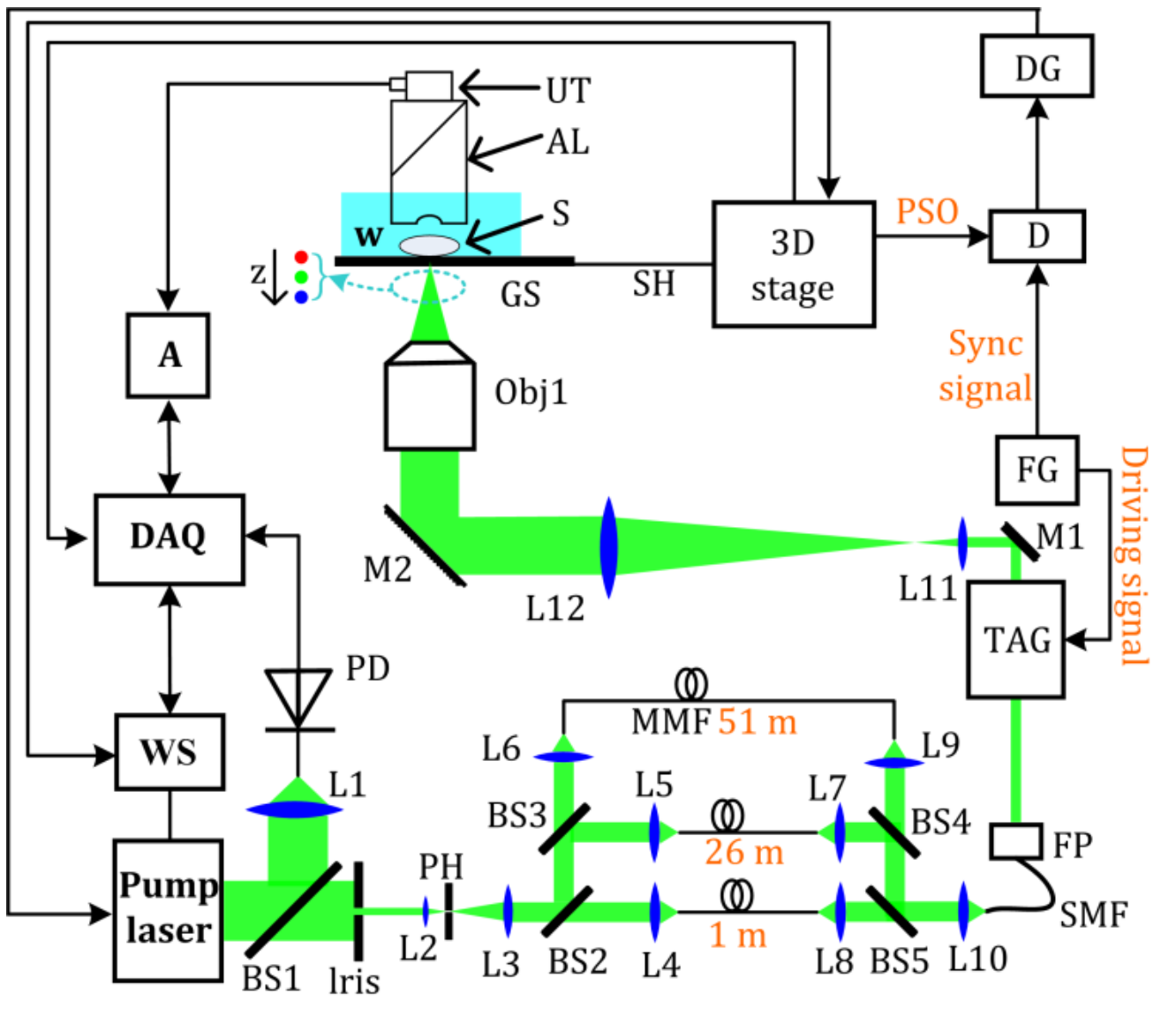

Fig. 1. Scheme of the varifocal OR-PAM system. A, amplifier; AL, acoustic lens; BS1, BS2, …, BS5, beam sampler; DAQ, data acquisition card; DG, Digital Delay and Pulse Generator ; D, D type flip-flop; FP, fiber port; FG, function generator; GS, glass slide; M1 and M2, mirrors; MMF, multimode fiber; L1, L2, …, L12, optical lenses; Obj1, objectives; PD, photodiode; PH, pinhole; S, sample; SH, sample holder; SMF, single mode fiber; TAG, TAG lens; UT, ultrasound transducer; W, water tank; WS, work station.

The light come from TAG is delivered into a beam expander formed by plano-convex lenses L11 (f = 18 mm) and L12 (f = 150 mm) with a magnification factor of 8.3. The TAG lens is conjugated to the back focal plane of the objective (5× Olympus objective, N. A. 0.14). A home-made acoustic lens (NA=0.5) is glued on a ultrasound transducer (central frequency 50MHz, V214-BB-RM, Olympus) to detect the ultrasound [19]. To couple the photoacoustic signals, we use a water tank. Both of sample and water tank are mounted on a 3D scanning stage which is assembled by a 2D linear stage (ANT95-XY, Aerotech) and a lifting stage (M-Z01.5G0, PhysikInstrumente). The 2D linear stage is used for the raster scanning of the sample. Photoacoustic signals detected by the ultrasound transducer is amplified by a amplifier (AU-1291, MITEQ) and acquired by a data acquisition card (ATS9350, Alazartech). A sinusoidal signal with a frequency of 707 kHz generated by a function generator (DS345, Stanford Research Systems) drives the TAG lens at an eigenmode. The Sync Output of the function generator, which provides the synchronizing TTL square wave, is connected to the clock input of a D-type flip-flop (SN74AUC1G74, Texas Instruments). Position synchronized output (PSO) signal of the 2D scanner is sent to the data input. Then the laser , scanner, and TAG lens are synchronized. A Digital Delay and Pulse Generator (DG535, Stanford Research Systems) delays the output signal from the flip-flop for laser triggering.

The home-made TAG lens used in our system consists of a cylindrical piezoelectric shell (PZT-8, Boston Piezo Optics) filled with a transparent silicone oil (100 cS, Sigma-Aldrich), with a refractive index of 1.403 and a speed of sound of 1000 m s−1. The length of cylindrical piezoelectric shell is 20 mm with Inner diameter of 16 mm and Outer diameter of 20 mm. The focusing power of the TAG lens exhibits a sinusoidal oscillation at the frequency of the driving signal [20]. Since the trigger signal of the laser is synchronized with TAG driving signal, different lag time of laser pulse relative to the TAG driving signal allows us to synchronize pulse of light with the desired vibration state of the lens. There are ~120 ns lag time between laser pulse out of 1 m- length MMF and 26 m- length MMF. Therefore, the laser pulse out of three fibers synchronize with three vibration states of the lens, respectively. And we finally achieved multifocus.

We then imaged a sharp bar edge of an USAF 1951 resolution target (Edmund Optics) indicated in Fig. 2(a) to estimate the lateral resolution with step size of 0.5 μm. During the test, there was no signal driven on the TAG lens. An edge spread function (ESF) profile was acquired by summing the acquired 3D data along the y-axis and then projecting along the depth direction by maximum amplitude projection (MAP), resulting in an ESF profile that was averaged 60 times. The lateral resolution is defined as the full width at half maximum (FWHM) of the line spread function (LSF) that is the derivative of the ESF, As shown in Fig. 2(b), the lateral resolution is estimated to 2.8 μm.

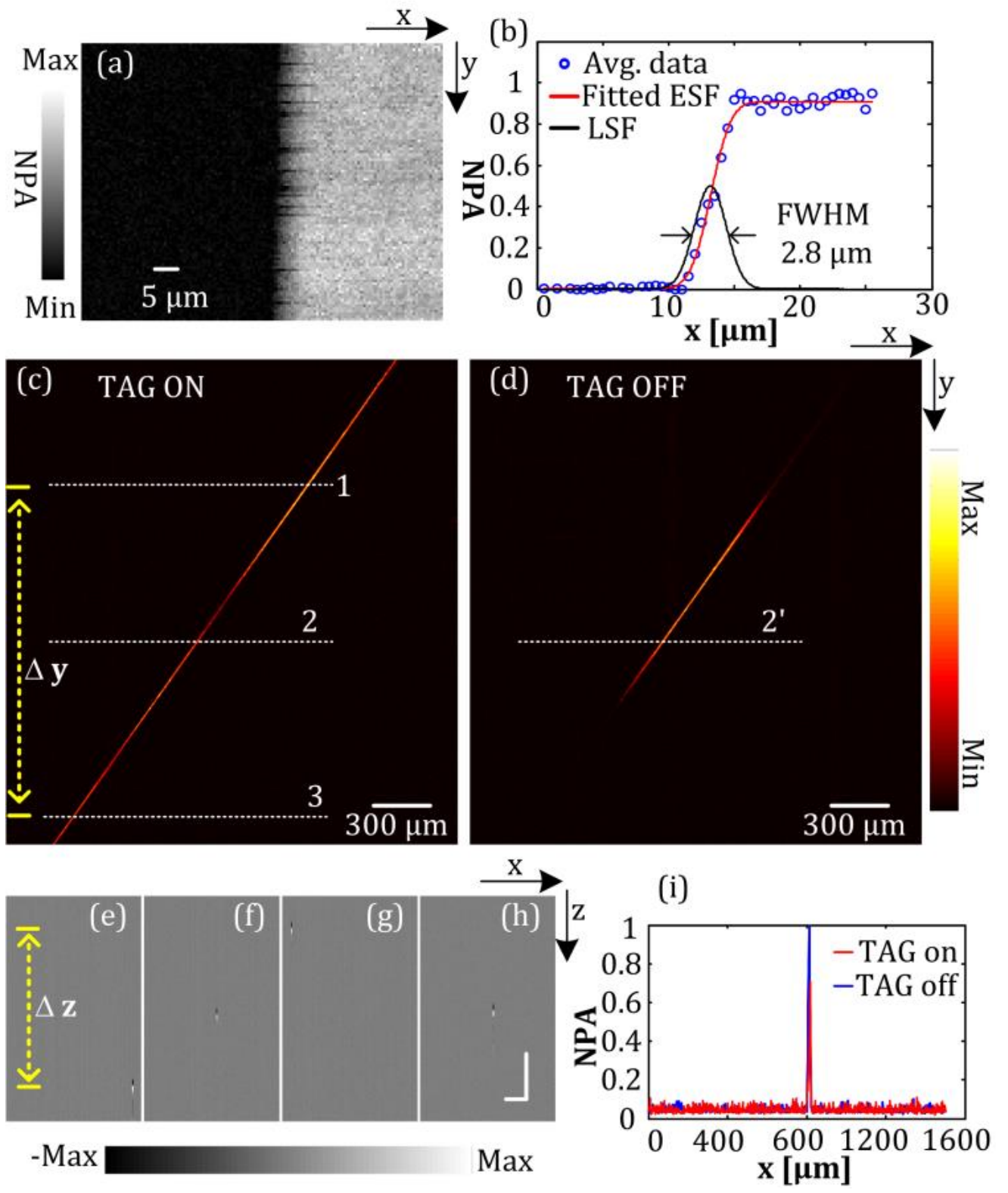

Fig. 2. (a) PA image of a bar edge on the resolution target. (b) Edge spread function (ESF) and line spread function (LSF) extracted from (a). (c) and (d) are MAP images of the vertically tilted carbon fiber when TAG lens on and off, respectively, Δ y = 1860 μm. (e), (f), (g) and (h), Close-up cross sectional B images through the dashed lines 1, 2 and 3 in (c) and 2' in (d), respectively, Δ z = 775 μm. (i) Profiles through the dashed lines 2 and 2'. NPA, normalized photoacoustic amplitude; Scale, 220 μm.

The DoF of the MF-PAM system is estimated by a vertically tilted carbon fiber with a diameter of ~6 μm. A small piece of glass was glued onto a glass slide to form a raised platform. One end of a straightened carbon fiber was fixed on the glass slide, while the other end was fixed to the platform. 2D raster scan was performed on the vertically tilted carbon fiber to acquire the depth-dependent PA signal distribution of the fiber. We imaged the vertically tilted carbon fiber using MF-PAM (TAG lens on, the driving signal is 20 $V_{p-p}$, 707 kHz), then TAG lens off (There was no signal driven on the TAG lens, only has a fix focus during the scanning) for compare. Fig. 2(c) and Fig. 2(d) are the MAP images that correspond to the TAG lens on and off, respectively. More information in axial can be acquired by MF-PAM, entire vertically tilted carbon fiber were mapped, whereas only partial can be reveal when TAG lens off. Fig. 2(e), 2(f), 2(g) are the close-up cross-sectional B images through the dashed lines 1, 2 and 3 in Fig. 2(c), respectively, and cover over a depth range of 775 μm in z-axis and 1860 μm in y-axis. Since the carbon fiber was straightened out, the depth variation was linear during the scanning, we can estimated the slope of the vertically tilted carbon fiber is 0.42. Fig. 2(h) is the close-up cross-sectional B images through the dashed line 2' in Fig. 2(d). The distribution of the PA amplitude along the dashed lines 2 and 2' are shown in Fig. 2(i), the width retain relatively consistent when TAG lens on and off. The carbon fiber was imaged over a range of 2400 μm in y-axis, we can estimated that the DoF is larger than 775 μm.

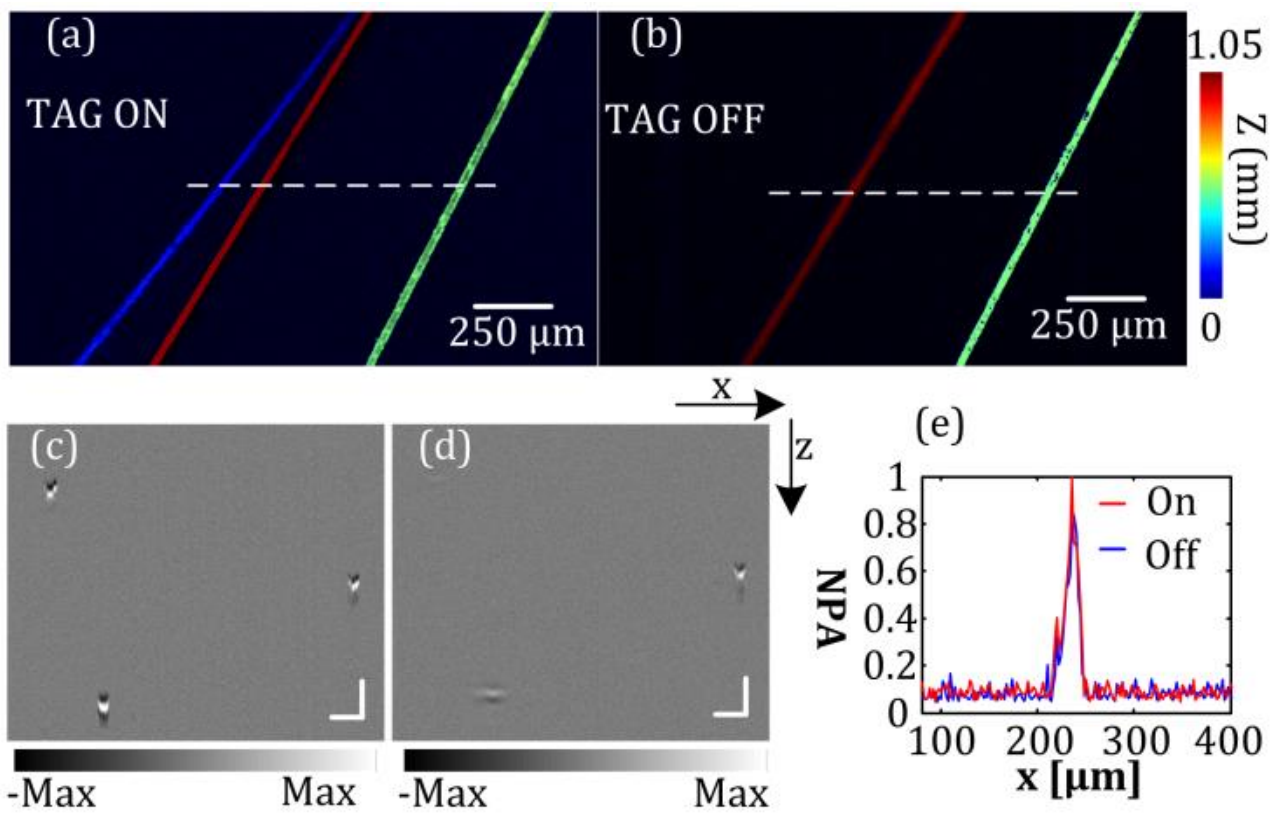

Fig. 3. Tungsten wire network imaged by MF-PAM. (a) and (b) are depth-coding MAP images of the tungsten wire network when TAG lens on and off, respectively. (c) and (d), Close-up cross sectional B images through the dashed line in (a) and (b). (e) Distribution of the PA amplitude of the middle tungsten wires along the dashed line in (a) and (b). NPA, normalized photoacoustic amplitude; Scale, 80 μm.

A phantom was imaged to demonstrate the extended DoF of MF-PAM system. The Phantom was composed of three tungsten wires (~20 μm diameter), a glass slide, and several fragments of cover glass (170 μm thick), the top and the middle tungsten wire was lifted ~680 μm and ~340 μm above the bottom tungsten wire, respectively, forming a network with a depth range of more than 700 μm. Figures 3(a) shows depth-coding MAP images obtained by MF-PAM, TAG lens on. Figures 3(b) shows depth-coding MAP images obtained when TAG lens off. Owing to the much larger DoF of MF-PAM, the bottom tungsten wire and the top tungsten wires could be revealed clearly, whereas only the middle tungsten wire could be reveal when the system has only a fix focus. Figures 3(c) and (d) are close-up cross sectional B images through the dashed line in Figures 3(a) and (b). The top two tungsten wires and the bottom two tungsten wires have a distance of ~400 μm and ~530 μm in depth, respectively, which are within the detection range of MF-PAM. However, in Fig. 3(d), only the middle tungsten wire is detected clearly. Since the bottom tungsten wire is out of detection range, it broadens seriously with week PA amplitude. Fig. 3(e) shows the profiles of the middle tungsten wires along the dashed line in (a) and (b).

We also showed that this system can be used for imaging of biological samples. A 30-days-old zebrafish (AB strain) was chosen for in- vivo imaging. Before the imaging, low melting point agarose (A-4018, Sigma-Aldrich) was dissolved at 40 °C (1.2% w/v), a culture dish was previously coated with a thin layer of the melted liquid agarose. When the temperature of liquid agarose dropped to 37 °C, a zebrafish was placed in the culture dish, lightly covered with the melted liquid agarose and oriented so that it was lying on its back waiting for polymerization. After the polymerization, some double-deionized water was pour into the culture dish. The temperature was kept around 25 °C during the imaging.

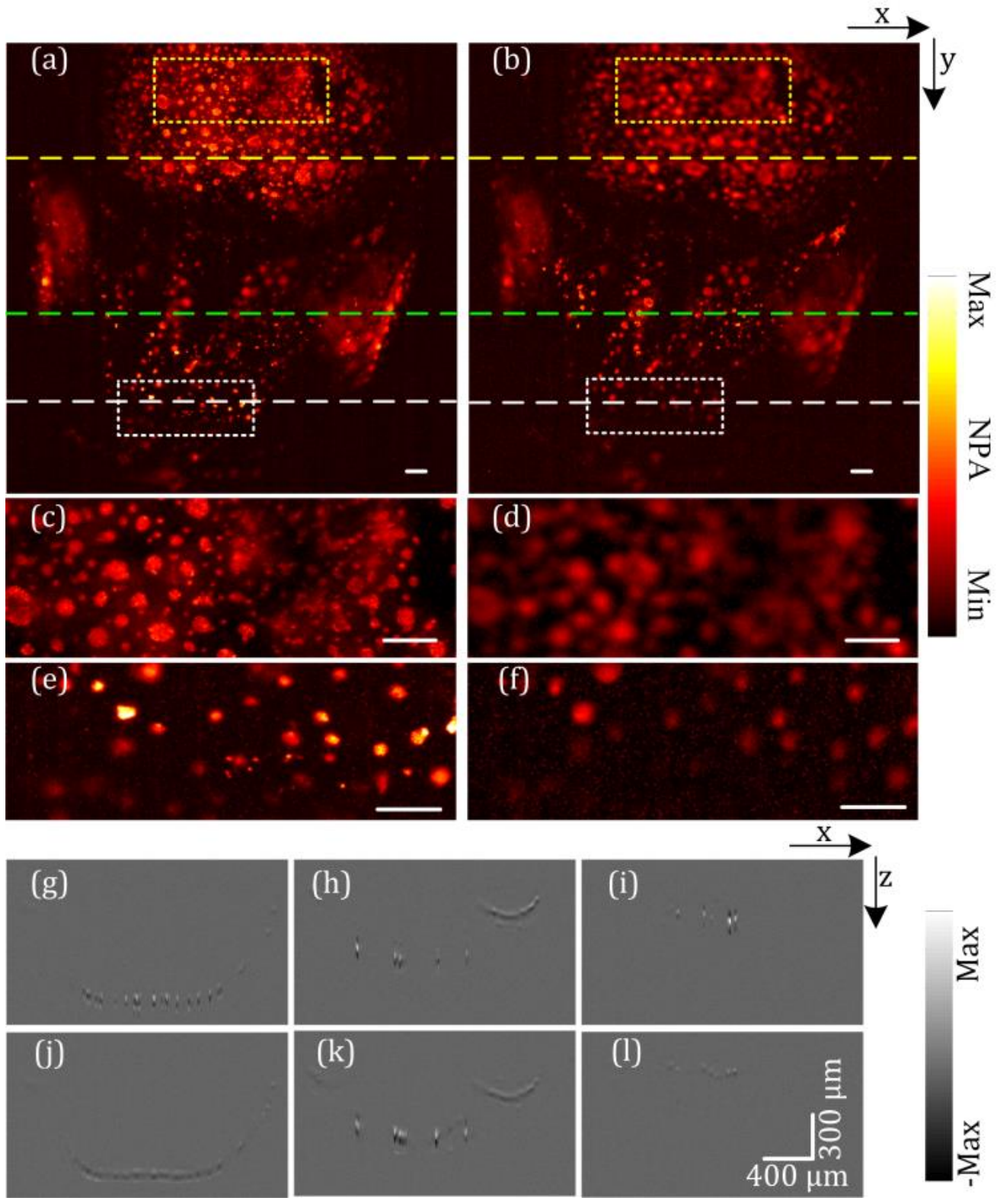

Fig. 4. Images of the head of a zebrafish. (a) and (b) are the MAP images of the head of a zebrafish when TAG lens on and off, respectively. (c) and (d) are close-up images of the areas indicated by the yellow dashed rectangles in (a) and (b), respectively; (e) and (f) are close-up images of the areas indicated by the white dashed rectangles in (a) and (b), respectively. (g)–(i) are cross-sectional B images through the yellow, green and white dashed lines in (a), respectively; (j)–(l) show cross-sectional B images through the yellow, green and white dashed lines in (b), respectively. NPA, normalized photoacoustic amplitude. Scale, 100 μm.

Figure 4 shows in vivo images of a zebrafish head. Figures 4(a) and (b) are MAP images when TAG lens on and off, respectively. Equipped with a much larger DoF, MF-PAM could aquire more details. Fig. 4(c) and (d) are close-up images of the areas indicated by the yellow dashed rectangles in Fig. 4(a) and (b), respectively; Fig.

4(e) and (f) are close-up images of the areas indicated by the white dashed rectangles in Fig. 4(a) and (b), respectively. The yellow and white dashed rectangles highlight the capability of MF-PAM to provide more structural details. The cross-sectional B images Fig. 4(g)- (l) also show the capability of MF-PAM to improve DoF. In Fig. 4(g)–(i) that are correspond to Fig. 4(a), the profile of pigments from bottom to top can be distinguished. However, In Fig. 4(j)–(l) that are correspond to Fig. 4(b), only middle part was clearly visualized , the others were quiet blurred.

In summary, by using a TAG lens and a fiber delay network into the PAM system, we developed a multifocus photoacoustic microscope. The TAG lens is used to high -speed focus-shift, and the fiber delay network output three laser beams with different delay time, and the laser pulse of each synchronizes with a vibration state of the lens, which we can obtain three focuses to improve the DoF of PAM system. In the experiments above, a 20 $V_{p\text{-}p}$ sinusoidal signal with frequency of 707 kHz was driven on the TAG lens when TAG lens on. The DoF we measured by a vertically tilted carbon fiber is eatimated to larger than 775 μm. The lateral resolution by imaging a sharp bar edge of an resolution target is 2.8 μm. The large DoF of MF-PAM was also verified by imaging a tungsten wire network and a zebrafish. It is worth noting that the MF-PAM improves the DoF without sacrificing the capability of functional imaging. The capability of MF-PAM system could be used for axial scanning, which may fulfill different requirements in biomedical researches.

**Funding.** National Science Foundation (NSF) (1263236, 0968895, 1102301); The 863 Program (2013AA014402)